\documentclass[12pt,a4paper]{article}
\usepackage{graphicx}
\usepackage[cp1251]{inputenc}
\usepackage{amssymb, amsfonts, amsmath}

\begin{document}

\title{
The Dirac equation in an external electromagnetic field: symmetry algebra and exact integration}

\author{A.I. Breev\footnotemark[1] \footnotemark[2]\ \ and A.V. Shapovalov\footnotemark[1] \footnotemark[2]}

\date{}

\renewcommand{\thefootnote}{\fnsymbol{footnote}}
\footnotetext[1]{Tomsk State University, Lenin avenue 36, Tomsk 634050, Russia.}
\footnotetext[2]{Tomsk Polytechnic University, Lenin avenue 30, Tomsk 634050, Russia.}

\maketitle

\begin{abstract}
	Integration of the Dirac equation with an external electromagnetic field is explored in the framework of the method of separation of variables and of the method of non\-com\-mu\-ta\-ti\-ve integration. We have found a new type of solutions that are not obtained by separation of variables for several external electromagnetic fields. We have considered an example of crossed electric and magnetic fields of a special type for which the Dirac equation admits a nonlocal symmetry operator.
\end{abstract}

\section*{Introduction}

The Dirac equation for a charge in an external electromagnetic field is the basic equation for relativistic quantum mechanics and quantum electrodynamics. In relativistic quantum mechanics the Dirac equation is interpreted as a single-particle wave equation describing fermions in an external field \cite{Bagrov, BagrovG,Simulik, Sitenko}. In quantum electrodynamics, exact solutions of the Dirac equation are needed to obtain the Furry interaction picture to keep exactly the interaction with the external field and the interaction of photons \cite{FrGitShv}.

To construct exact solutions of the Dirac equation, the separation of variables method (SoV) is commonly used. The external electromagnetic fields admitting SoV in the Dirac and the Klein--Gordon equations have been listed in refs \cite{Bagrov,BagrovG}.

In refs. \cite{SpSh1,SpSh2} a new method, named the noncommutative integration (NI) method, has been proposed to construct basises of exact solutions of linear partial differential equations. The NI method essentially uses a Lie algebra $\mathfrak{g}$ of differential symmetry operators of the first order.

Integration of the free Dirac equation using the NI method was considered in refs. \cite{Br01,Br02,Klish1}. Note that the method allows one to find exact solutions (NI-solutions) in the cases when the Dirac equation does not allow separation of variables \cite{Klish2, Varaksin}.

The paper is organized as follows. In the next section we introduce basic notations used below and briefly describe the SoV method and the NI method applied to the Dirac equation. In Section 2 we apply the NI method for the Dirac equation in a spherically symmetric electromagnetic field. In section 3  we study the case when the magnetic field is uniform and constant. In section 4 we consider integration of the Dirac equation with the electromagnetic field which contracts the symmetry algebra of the free Dirac equation to a commutative subalgebra. Concluding remarks are in Section 5.

\section{Symmetry algebra  of the Dirac equation and construction of exact solutions}

Let us consider the four-dimensional Lorentz manifold $M$ with coordinates $x^\mu$ $(\mu,\nu = 1,\dots,4)$ having the line element $ds^2=g_{\mu\nu} (x) dx^\mu dx^\nu$. The Dirac equation for a spinor $\psi$ on $M$ can be written as\footnote{We will use a unit system in which $\hbar=c=1$ and the Minkowski metric is $\eta_{\mu\nu} =\operatorname*{diag} (1, -1, -1, -1)$}
\begin{equation}
	\hat H \psi = m\psi,\quad \hat H = \gamma^\mu \hat{\mathcal{P}}_\mu,\quad
	\hat{\mathcal{P}}_\mu = \hat p_\mu - e A_\mu(x),\quad \hat{p}_\mu = i(\partial_\mu + \Gamma_\mu),\quad x\in M.
	\label{dirac_eq}
\end{equation}
Here, $A_\mu(x)$ is a potential of an external electromagnetic field, $e$ and $m$ are charge and mass of a particle, respectively. The Dirac matrices $\gamma^\mu$ are  defined as a solution of the system $\{\gamma^\mu, \gamma^\nu \} = 2 g^{\mu \nu} (x) $. The spin connection $\Gamma_\mu$ is defined by \cite{Klish1}
\begin{equation}\nonumber
	\Gamma_\mu = -\frac{1}{4}\gamma_{\nu;\mu}\gamma^{\nu},\quad
	\gamma_{\nu;\mu} = \partial_{\mu}\gamma_\nu - \Gamma^\rho_{\nu\mu}\gamma_\rho,
\end{equation}
where $\Gamma^\rho_{\nu \mu}$ are the  Christoffel symbols of the symmetric connection compatible with the metric $g_{\mu \nu} (x)$.

The matrices
\begin{gather}\nonumber
	E_4,\quad \gamma^\mu,\quad
	\gamma^{\mu\nu} = \frac{1}{2}[\gamma^\mu,\gamma^\nu],\quad
	\gamma = -\frac{1}{4!}e_{\mu\nu\tau\rho}\gamma^\mu\gamma^\nu\gamma^\tau\gamma^\nu,\quad
	\stackrel{*}{\gamma}_\mu = -\frac{1}{3!}e_{\mu\nu\tau\rho}\gamma^\nu\gamma^\tau\gamma^\rho,
\end{gather}
form a basis for $4\times 4$ matrices. Here $e_{\mu\nu\tau\rho} = \sqrt{-\operatorname{det}(g_{\mu\nu}(x)}\varepsilon_{\mu\nu\tau\rho}$ is totally antisymmetric tensor,
($\varepsilon_{1234} = 1$).

In Minkowski space $\mathcal{M}$ for the Dirac gamma matrices we use the standard representation
\begin{equation}\nonumber
   \gamma^1 = \begin{pmatrix} 1 & 0 \\ 0 & -1 \end{pmatrix};\quad
   \gamma^{k+1} = \begin{pmatrix} 0 & \sigma^k \\ -\sigma^k & 0 \end{pmatrix},\quad
   \Sigma^k = \begin{pmatrix} \sigma^k & 0 \\ 0 & \sigma^k \end{pmatrix},\quad k = 1,2,3,
  \end{equation}
where $\sigma^k$  are the usual Pauli matrices.

The matrix differential symmetry operators of the first order for the Dirac equation (\ref{dirac_eq}) have been obtained as a linear combination of the following three independent operators (see \cite{ShVN} and \cite{Carter}):
\begin{gather}
	\label{killX}
	\hat X = \xi^\mu(x)\hat{\mathcal{P}}_\mu - \frac{i}{4}\gamma^{\mu\nu}\xi_{\mu;\nu}(x) + \varphi(x),\\
	\label{YakillL}
	\hat L = \stackrel{*}{\gamma}_\mu f^{\nu\mu}(x)\hat{\mathcal{P}}_\nu +\frac{i}{3}\gamma_\mu\tilde{f}^{\nu\mu}_{;\nu}(x),\\
	\label{YaJ}
	\hat J = 2\gamma\gamma^{\mu\nu}f_\nu(x)\hat{\mathcal{P}}_\mu + \frac{3 i}{4}\gamma f^\mu_{;\mu}(x),
\end{gather}
where $\tilde{f}_{\mu\nu} (x) = \frac{1}{2}e_{\mu\nu\tau\rho} f^{\tau \rho}(x)$ is the dual tensor of the tensor $f^{\mu\nu}(x)$.
We call operators (\ref{killX}) \textit{the Killing symmetry operators} as they are expressed in terms of the Killing vector field $\xi^\mu (x)$,
\begin{equation}\nonumber
	\xi_{\mu;\nu}(x) + \xi_{\nu;\mu}(x) = 0, \quad
	\partial_\nu\varphi(x) = F_{\nu\mu}(x)\xi^\mu(x).
\end{equation}
Here $F_{\nu\mu}(x) = \partial_\nu A_\mu (x) - \partial_\mu A_\nu (x) $ is the tensor of an electromagnetic field. The symmetry operators (\ref{YakillL}) are given by the antisymmetric Yano-Killing tensor field $f_{\mu\nu}(x)$:
\begin{equation}\nonumber
	f_{\mu\nu;\tau}(x) + f_{\mu\tau;\nu}(x) = 0,\quad
	{f_\nu}^\alpha(x) F_{\alpha\mu}(x) = {f_\mu}^\alpha(x) F_{\alpha\nu}(x).
\end{equation}
The symmetry operations (\ref{YaJ}) are determined in terms of the Yano vector field $ f_\mu (x) $:
\begin{equation}\nonumber
	f_{\mu;\nu}(x) = \frac{1}{4}g_{\mu\nu}(x){f^\mu}_{;\mu}(x),\quad
	F_{\nu\mu}(x)f^\mu(x) = 0.
\end{equation}
We call operators (\ref{YakillL}) and  (\ref{YaJ}) \textit{ the spin symmetry operators}.

A necessary condition for the SoV in the Dirac equation is the existence of three mutually commuting independent symmetry operators of the first order. Under this condition, the problem of finding solutions of the Dirac equation is reduced to the eigenvalue problem
\begin{equation}
\hat H \psi = m \psi,\quad {\hat S}_a \psi = \lambda_a \psi,\quad a = 1,\dots,3,
	\label{RPzad}
\end{equation}
where ${\hat S}_a$ are symmetry operators of the form (\ref{killX}) -- (\ref{YaJ}), and $\lambda_a$ are essential separation parameters.

The separation of variables method can be applied to the squared Dirac equation \cite{Obukhov} which is obtained from the Dirac equation (\ref{dirac_eq}) by  the substitution $\psi = (\hat H + m) \varphi$. Then for $\varphi $ we obtain \textit{the squared Dirac equation} reads $(\hat {H}^2 - m^2)\varphi = 0$. If there exists a constant matrix $\Omega$ such that $(\hat{H}^2 - m^2) \Omega = \Omega\hat {L}_2$ where $\hat{L}_2$ is a diagonal second order differential operator, we say that the squared Dirac equation \textit{admits diagonalization}. Then the complete system of solutions of the Dirac equation can be written as $\psi = (\hat H + m)\Omega\Phi$ where $\Phi$ is a complete system of solutions for the equation $\hat{L}_2\Phi = 0 $. The last one is integrated in the framework of the SoV method.

Let $\mathfrak{g}$ be a Lie algebra of the Killing symmetry operators (\ref{killX}). Denote by $G = \exp(\mathfrak{g})$ a Lie group of motions on $M$. The spaces $\mathfrak{g}^*$ and $\mathfrak{g}$ are conjugates of each other. Consider a neighborhood  $U_0$ of a regular point $x_0 \in M$ in which we will seek a basis of solutions of the Dirac equation. In this neighborhood we introduce a new coordinate system $(r, z)$ where $r$ are independent invariants of a Lie group $G$, and $z$ are local coordinates on an orbit of $x_0$. Denote by $\mathfrak{h}$ a stationary Lie algebra of the point $z_0$. Without loss of generality, we assume that the Killing symmetry operators depend only on the coordinates $z$. The method consists in generalization of the eigenvalue problem (\ref {RPzad}) as follows \cite{SpSh1}:
\begin{equation}
	\hat X_a(z) \psi(x; q,\lambda) = - l_a(q,\partial_q,\lambda) \psi(x; q,\lambda),\quad a = 1,\dots,\operatorname*{dim}\mathfrak{g},
	\label{sysXl}
\end{equation}
where $l_a(q,\partial_q,\lambda) = \alpha^\nu_a(q)\partial_{q^\nu} + i \chi_a(q,\lambda)$ are operators of an irreducible
$ \lambda$-representation of the algebra $\mathfrak{g}$ constructed using a K-orbit $\mathcal {O}_\lambda$ passing through the covector $\lambda\in\mathfrak{g}^*$. The variables $q$ are local coordinates on a Lagrangian submanifold $Q$ of the orbit $\mathcal {O}_\lambda $. If the Lie algebra $\mathfrak {g}$ is Abelian, then $l_a (q, \lambda) = \lambda_a$ and we come to the SoV method. The basis of solutions of equation (\ref{sysXl}) is represented as $\psi(x; q, \lambda) = \exp(R(x; q, \lambda)) \hat \psi(u; r, \lambda)$, where $R(x; q, \lambda)$ is a certain function, and $\hat\psi(u; r, \lambda)$ is an arbitrary function depending on the characteristics of the system $u = u(q, x)$. Substituting this solution into the Dirac equation (\ref{dirac_eq}), we obtain the reduced equation

\begin{equation}
	\hat H(u,\partial_u, r, \partial_r; \lambda)\hat \psi(u; r, \lambda) = m\hat \psi(u; r,\lambda).
	\label{redoH}
\end{equation}
The number of independent variables of the reduced equation (\ref{redoH}) is shown to be determined by algebraic properties of the orbit of $ z_0 $ \cite{Br14}.

If equation (\ref {redoH}) is algebraic or an ordinary differential equation, it is said that the original Dirac equation (\ref{dirac_eq}) is \textit{noncommutatively integrable}.
Note that the functions $\psi (z; q, \lambda) $ are the eigenfunctions of all nontrivial Casimir operators
\begin{equation}\nonumber
	\hat K(z,\partial_z)\psi(z; q,\lambda) = \kappa(\lambda)\psi(z; q,\lambda).
\end{equation}
The basis of solutions of equation (\ref{dirac_eq}) can be represented as a generalized Fourier transform
\begin{equation}\nonumber
	\psi(z; q,\lambda) = \int_U \psi(u; r, \lambda) D^\lambda_{q u}(x) d\mu(u),
\end{equation}
where the distributions  $D^\lambda_{qu} (x) $ have the properties of completeness and orthogonality \cite{Br14, ShSO3}.

Consider the important case in which the Lie group $G$ acts on $M$ simply transitively and the neighborhood $U_0$ can be identified with a local Lie group $G$ by entering the group coordinates $g = g(z)$. Without loss of generality, we assume that the group acts by right translations. The noncommutative integration method in this case may be formulated using Kirillov's orbit  method as follows \cite{ShSO3, Br02}. The basis of solutions of equation (\ref{dirac_eq}) reads
\begin{equation}\nonumber
	\psi(g; q, \lambda) = \int_Q \psi(q'; q,\lambda) D^\lambda_{q q'}(g^{-1}) d\mu(q'),\quad
	q,q' \in Q,
\end{equation}
where $D^\lambda_{q q '}(g) $ are generalized kernels of an irreducible unitary representation of the Lie group $G$ constructed on a non-degenerate K-orbit $\mathcal {O}_\lambda$ (see \cite{Kirr}). The representation of the Lie algebra $\mathfrak{g}$ is at the same time the $\lambda$-representation of $\mathfrak{g}$. The distributions $D^\lambda_{q q'}(g)$ can be found from the following system of equations:
\begin{equation}
	\left(\xi_a(g) + \overline{l_a(q,\partial_q)}\right) D^{\lambda}_{q q'}(g) = 0,\quad
	\left(\eta_a(g) + l_a(q',\partial_{q'})\right)D^{\lambda}_{q q'}(g) = 0.
	\label{sysXLg}
\end{equation}
Here $\xi_a(g) $ and $\eta_a(g) $ are the left-invariant and right-invariant vector fields on $U_0$, respectively. The functions $\psi(q',\lambda) $ satisfy the reduced equation
\begin{equation}\nonumber
	\hat H(q',\partial_{q'}; q, \lambda)\hat \psi(q'; q, \lambda) = m \psi(q'; q,\lambda)
\end{equation}
with the number of independent variables $q'$ equal to $\operatorname{dim}Q = (\operatorname{dim}\mathfrak{g} - \operatorname{ind}\mathfrak{g})/2$. Here $\operatorname{ind}\mathfrak{g}$ is the index of the Lie algebra $\mathfrak{g}$ defined as the dimension of the annihilator of a generic covector.

\section{The Dirac equation in a spherically symmetric electric field}

Let us define a potential $A_\mu(x)$ by the relations
\begin{equation}\nonumber
	A_1 = V(r),\quad A_k = 0,\quad r = \sqrt{x^2 + y^2 + z^2},\quad k=2,\dots,4,
\end{equation}
where $V(r)$ is a smooth function. The Dirac equation on $\mathcal{M}$ can be written in a Hamiltonian form as $i\partial_t\psi = \hat H \psi$. The Hamiltonian $\hat{H}$ in spherical coordinate system $(r,\theta,\phi)$ takes the form:
\begin{equation}\nonumber
	\hat{H} = -i\alpha^1\left(\frac{1}{r} +\partial_r \right)-i\alpha^2 \frac{1}{r}\left(\frac{1}{2}\cot{\theta}+\partial_\theta \right) - i\alpha^3\frac{1}{r\sin{\theta}}\partial_\phi + \beta m + e V(r),
\end{equation}
where 
$\beta = \gamma^1, \alpha^1 = \gamma^1\gamma^2,\ \alpha^2 = \gamma^1\gamma^3,\ \alpha^3 = \gamma^1\gamma^4$.
Consider the stationary Dirac equation
\begin{equation}\label{statDirac}
   \hat H\psi = E\psi,\quad \psi = \psi(r,\theta,\phi).
\end{equation}
The Killing symmetry operators corresponding to the generators of the rotation group $ SO (3) $ are given by
\begin{gather}\nonumber
	\hat X_1(\theta,\phi) = \partial_\phi,\quad
	\hat X_2(\theta,\phi) = -\cot\theta\sin\phi\partial_\phi+
	      \cos\phi \partial_\theta+\frac{i}{2}\frac{\sin\phi}{\sin\theta}\Sigma^1,\\ \nonumber
	\hat X_3(\theta,\phi) = -\cot\theta\cos\phi\partial_\phi-
  	       \sin\phi\partial_\theta+\frac{i}{2}\frac{\cos\phi}{\sin\theta}\Sigma^1.
\end{gather}
These operators the following realization of the Lie algebra $\mathfrak{so}(3)$:
\begin{equation}\nonumber
	[\hat X_1,\hat X_2] = \hat X_3,\quad [\hat X_3,\hat X_1] = \hat X_2,\quad [\hat X_2,\hat X_3] = \hat X_1.
\end{equation}
Note that inclusion of the external electromagnetic field does not affect the algebra of symmetry operators $\hat X_a(\theta, \phi)$. The Casimir operator of the algebra $\mathfrak {so}(3)$ is the square of the total angular momentum:
\begin{equation}\nonumber
	\mathbf{J}^2 = K(-i\hat X), \quad K(f) = f_1^2 + f_2^2 + f_3^2.
\end{equation}
The spin Yano-Killing symmetry operator reads
\begin{equation}\nonumber
	\hat S  = -\beta \left( \Sigma^2\frac{1}{\sin\theta}\partial_\phi - \Sigma^3 \left[\frac{1}{2}\cot\theta+\partial_\theta\right]\right),\quad [\hat S, X_a] = 0.
\end{equation}
Consider briefly the separation of variables in equation (\ref{statDirac}) (for more details see \cite{Bagrov} and \cite{Akhiezer}).
There exists a complete set $\{\hat S, -i X_3, \mathbf {J}^2\}$ of mutually commuting symmetry operators. The solution of the eigenvalue problem
\begin{gather}\label{z1}
	\mathbf{J}^2\psi_{j M \zeta} = j(j+1)\psi_{j M \zeta},\quad
	\hat S \psi_{j M \zeta} = \zeta \left( j+\frac{1}{2} \right)\psi_{j M \zeta},\quad
	-i X_3 \psi_{j M \zeta} = M \psi_{j M \zeta}
\end{gather}
is obtained as
\begin{gather}\label{r1}
	\psi_{j M \zeta}(r,\theta,\phi) = \frac{1}{r}
	\begin{pmatrix} \Omega^j_{M \zeta}(\theta,\phi) f(r) \\
	          -\zeta\Omega^j_{M -\zeta}(\theta,\phi) g(r),
	\end{pmatrix},\quad
	M = -j\dots j,\quad \zeta = \pm 1,\quad j = 1,2,3,\dots,
\end{gather}
where $\Omega^j_{M\zeta}(\theta, \phi)$ is a spherical spinor \cite{Edmonds}. The Dirac equation (\ref{statDirac}) is reduced to a radial stationary Dirac equation with a fixed angular momentum $j$ and a projection of angular momentum in the direction of  $M$, and a spin quantum number $\zeta $. The radial Dirac equation for $\psi_{jM\zeta}$ is a system of ordinary differential equations for the functions $f(r)$ and $g(r)$:
\begin{gather}\label{radeq}
	f'(r)- \frac{1}{r}\zeta \left(j+\frac{1}{2}\right) f(r) - (E+m-e V(r)) g(r) = 0,\\ \nonumber
	g'(r)+ \frac{1}{r}\zeta \left(j+\frac{1}{2}\right) g(r) + (E-m-e V(r)) f(r) = 0.
\end{gather}
Let us perform the noncommutative reduction of the Dirac equation (\ref{statDirac}) with the use of the Lie algebra $\mathfrak{so}(3)$ of symmetry operators of the first order $\hat X_a(\theta, \phi)$. Denote an invariant of the  Lie group $SO(3)$ as  $r$ and let  $ (\theta, \phi)$ be  coordinates on an orbit of the Lie group $SO(3)$.
The noncommutative integration consists in solving the system of equations
\begin{equation}\label{noncommsys}
	X_a(\theta,\phi) \psi = - l_a(q,\lambda) \psi,\quad \hat H \psi = E \psi,
\end{equation}
where $l_a(q,\lambda)$ are operators of a $\lambda $- representation of the Lie algebra $\mathfrak{so}(3)$. The $\lambda $- representation is constructed for functions defined on a Lagrangian submanifold $Q$ of orbits of the coadjoint representation (K-orbits) of the Lie algebra $\mathfrak{so}(3)$ \cite{ShSO3, GnSO3}.
The manifold  $Q$ has the topology of the cylinder: $\operatorname{Re}q \in [0, 2\pi),\ \operatorname{Im}q\in\mathbb{R}^1$. A covector $\lambda=(j,0,0) \in \mathfrak {so}^{*}(3) ,\ j \in \mathbb {Z}$ parametrizes a nondegenerate K-orbit. The operators $-il_a(q, \lambda)$ are Hermitian with respect to the scalar product
\begin{equation}\nonumber
	(\psi_1(q),\psi_2(q)) = \int_Q \overline{\psi_1(q)}\psi_2(q)d\mu(q),\quad
	d\mu(q) = \frac{(2 j + 1)!}{2^{j}(j !)^2}\frac{dq\wedge d\overline{q}}{(1+\cos(q-\overline{q}))^{j + 1}}
\end{equation}
and have the form
\begin{gather}\nonumber
	l_1(q,\lambda) = -i(\sin(q)\partial_q - j \cos(q)),\quad
	l_2(q,\lambda) = -i(\cos(q)\partial_q + j \sin(q)),\quad
	l_3(q,\lambda) = \partial_q.
	\label{lprSO3}
\end{gather}
Integrating  equations (\ref{noncommsys}), we obtain the basis of solutions
\begin{equation}\label{nonc1}
	\psi_\sigma(r,\theta,\phi) = \frac{1}{r}
	\begin{pmatrix}
		D^j_{q\zeta}(\theta,\phi) f_\sigma(r) \\
		-\zeta D^j_{q (-\zeta)}(\theta,\phi) g_\sigma(r)
	\end{pmatrix},\quad \sigma = (q,j,E,\zeta),
\end{equation}
where
\begin{gather}\label{nD1}
	D^j_q(\theta,\phi) = \frac{2^j(j!)^2}{(2j)!}  \bigg{(} -i\cos\theta\sin q + \sin\theta (\cos\phi-i\cos q\sin\phi) \bigg{)}^{j-1/2}\times \\ \nonumber
   \times \sqrt{\cos\theta\sin q+i\cos\phi(\cos q+\sin\theta)+\sin\phi+\cos q\sin\theta\sin\phi}\bigg{[}\\ \nonumber
   \frac{\cos q\cos\theta+i\sin q(\cos\phi+i\sin\theta\sin\phi)}{1+\cos q\sin\theta+\cos\theta\sin q\sin\phi}\sigma^1 + 1\bigg{]}
   \begin{pmatrix}
	   -i\zeta \\
	   1
   \end{pmatrix},
\end{gather}
and the functions $ f_\sigma(r)$ and $ g_\sigma (r)$ satisfy the radial equation (\ref{radeq}). Solutions (\ref{nonc1}) are the eigenvalues of the Casimir operator  $\mathbf {J}^2 $ and of the spin operator $\hat S$, but   (\ref{nonc1})  are not eigenfunctions for the operator of angular momentum $-i X_3$:
\begin{equation}
	\mathbf{J}^2\psi_\sigma = j(j+1)\psi_\sigma,\quad
	\hat S\psi_\sigma = \zeta\left( j+\frac{1}{2} \right)\psi_\sigma,\quad
	-i X_3 \psi_\sigma = i\partial_q \psi_\sigma.
	\label{sys33}
\end{equation}
The Fourier transform of functions (\ref{nD1}) are the spherical spinors:
\begin{equation}\label{omega2}
	\Omega^j_{M \zeta}(\theta,\phi) = \int_Q D^j_{q \zeta}(\theta,\phi) e^{i M q} d\mu(q).
\end{equation}
Indeed, given (\ref{sys33}), we obtain that (\ref{r1}) is a family of solutions for the Dirac equation (\ref{statDirac}) and  (\ref{r1}) satisfies equations (\ref{z1}) if the spherical spinors are defined by expression (\ref{omega2}). Thus, formula (\ref {omega2}) establishes a link between the basis of the solutions obtained using separation of variables and by the method of noncommutative integration. Solution (\ref{nonc1}) describes a state with  definite value of the orbital angular momentum $j$ and spin $\zeta$. Relation (\ref{omega2}) allows considering a complex parameter $q$ and a projection of the moment $M$ as dual variables.

Note that (\ref{nD1}) is expressed in terms of elementary functions, while the spherical spinors are expressed in terms of special functions.

\section{Electromagnetic field extending a subalgebra of symmetry operators}

Consider an electromagnetic field with nonzero components $E_z=V'(z),\ H_z=H=const $ in Cartesian coordinates $(x,y,z)$ and choose a potential in the form
\begin{equation}
	A_0 = -V(z),\quad \mathbf{A} = (0, H x, 0 ).
	\label{1122}
\end{equation}
The Dirac equation on $\mathcal{M}(t,x,y,z)$ with the potential (\ref{1122}) reads
\begin{equation}\label{Hdirac}
	i\partial_t\psi = \hat H\psi,\quad
	\hat H = -i\alpha^1 \partial_x-\alpha^2(i\partial_y+e H x)-i\alpha^3\partial_z - e V(z) + \beta m.
\end{equation}
The Killing symmetry operators can be written as
\begin{gather}\label{opXH}
	\hat X_0 = i e,\quad
	\hat X_1 = \partial_x - i e H y,\quad
	\hat X_2 = \partial_y,\quad
	\hat X_3 = y\partial_x - x \partial_y -\frac{i}{2}\Sigma^3 +
	i \frac{e H}{2}(x^2-y^2).
\end{gather}
Operators (\ref{opXH}) form a basis of the Lie algebra $\mathfrak{g}$ with non-trivial brackets:
\begin{equation}\nonumber
	[\hat X_1,\hat X_3] = -\hat X_2,\quad [\hat X_2,\hat X_3] = \hat X_1,\quad [\hat X_1,\hat X_2] = H \hat X_0.
\end{equation}
In the absence of the electromagnetic field, the symmetry operators (\ref{opXH}) form a Lie algebra $ \mathfrak {e} (2) $ of the Lie group $E(2)$ acting on the plane $(x, y) $. Inclusion of the external field leads to the fact that the Lie algebra of symmetry operators (\ref{opXH}) becomes a central extension of the Lie algebra $\mathfrak{e}(2)$.

We can show that if the symmetry operators of the first order without matrix coefficients at the derivatives form a Lie algebra, then this algebra will be a central extension of the Lie algebra of symmetry operators for the free Dirac equation \cite{Magazev}.

Note that in addition to (\ref{opXH}), there exists an additional spin symmetry operator
\begin{equation}
	\hat S = \beta \left( \Sigma^2\partial_x -
	\Sigma^1 (\partial_y - i e H x)\right),\quad [\hat S, \hat X_a] = 0.
	\label{spinH}
\end{equation}

Separation of variables in the Dirac equation is possible if we have a set of operators $ \{\hat H, \hat X_2, \hat S \}$ (see \cite{BagrovG, Akhiezer}). The basis of solutions decreasing at $|\xi| \rightarrow \infty $ has the form \cite{Akhiezer}:
\begin{gather}\label{solrazdH}
	\psi_\sigma(x,y,z) = e^{i p y}\begin{pmatrix}
		-i\frac{\zeta \sqrt{2}}{n} D_k(\xi) f(z)\\
		D_{k-1}(\xi) f(z)\\
		-i\frac{\zeta\sqrt{2}}{n} D_k(\xi)g(z)\\
		-D_{k-1}(\xi) g(z)
	\end{pmatrix},\quad \xi = \sqrt{\frac{2}{e H}}(e H x - p), \quad k = -n^2/2,
\end{gather}
where $D_k (\xi)$ are  parabolic cylinder functions. A spinor (\ref{solrazdH}) is a solution of the eigenvalue problem
\begin{equation}\nonumber
	-i\hat X_2\psi_\sigma = p\psi_\sigma,\quad
	\hat S\psi_\sigma = \zeta n \sqrt{e H} \psi_\sigma,\quad \sigma=(n,p,\zeta)\quad n = 0,1,2,\dots, \quad \zeta=\pm 1.
\end{equation}
The functions $f(z)$ and $g(z)$ satisfy the system of ordinary differential equations
\begin{gather}\label{reductH}
	 i f'(z) -  n\sqrt{e H}\zeta f(z) + (m + E + e V(z))g(z) = 0,\\ \nonumber
	i g'(z) + n\sqrt{e H}\zeta g(z) - (m - E - e V(z))f(z)  = 0.
\end{gather}
Consider the process of noncommutative reduction. The operators of $\lambda $ representation of the Lie algebra $\mathfrak{g}$ can be taken as follows:
\begin{equation}
	l_0 = -i e, \quad l_1 = -i \left( \frac{1}{2}\frac{\partial}{\partial q} - e H q  \right),\quad
	l_2 = \frac{1}{2}\frac{\partial}{\partial q} + e H q,\quad
	l_3 = - i q \frac{\partial}{\partial q},\quad q\in\mathbb{C}.
	\label{Lh}
\end{equation}
Operators (\ref{Lh}) are skew-Hermitian with respect to the measure
\[
d\mu(q) = \exp(2 e H |q|^2 )dq\wedge d\overline{q}.
\]
From the system $\hat X_a\psi = -l_a\psi$  we have
\begin{gather}\label{anzatzH}
	\psi(x,y,z) = \exp\left( e H \left[ \frac{x^2 +2 i x y + y^2}{4} - i q (x - i y) \right]\right)\left( \cosh(q') + \Sigma_3 \sinh(q') \right)\Phi(z),
\end{gather}
where $ q'= - \frac{1}{2}\log\left (q + \frac {i} {2} (x + iy) \right) $. Let us substitute (\ref {anzatzH}) into the equation $\hat H \psi = E \psi $, then we obtain the reduced equation in the form
\begin{gather}\nonumber
	\hat H' = \left( \frac{1}{4} - e H \right)\alpha^1 + i\left( \frac{1}{4} + e H \right)\alpha^2 -
	i \alpha^3\frac{d}{d z} - e V(z) + \beta m.
\end{gather}
Let us write the eigenvalue problem $\hat S\psi = \zeta\psi $ as follows:
\begin{gather}\nonumber
	\hat S'\Phi(z) = \zeta \sqrt{e H} \Phi(z),\quad
	\hat S' = \beta\left( e H (\Sigma^1 - i\Sigma^2) + \frac{1}{4}(\Sigma^1 + i\Sigma^2)\right).
\end{gather}
Then, we find the basis of solutions  parameterized by the spin quantum number $ \zeta $ and a complex parameter $ q $ for the
Dirac equation (\ref{Hdirac}):
\begin{equation}\label{noncH}
	\psi_\sigma(x,y,z) = \begin{pmatrix}
		D_{q,\zeta}(x,y) f_\sigma(z)\\
		D_{q,-\zeta}(x,y) g_\sigma(z)
	\end{pmatrix},\quad \sigma = (q,\zeta,E).	
\end{equation}
Here  the following notations are used:
\begin{gather}\nonumber
	D_{q,\zeta}(x,y) =  \exp\left( e H \left[ \frac{1}{4}(x+i y)^2 - i q (x - i y) +\frac{1}{2}y^2\right]\right)
	\begin{pmatrix} \exp(q') \\ 2 \exp(-q') \sqrt{e H}\zeta \end{pmatrix}.
\end{gather}
The functions $ f(z)$ and $g (z)$ are determined by the system of ordinary differential equations (\ref{reductH}) for $ n = 1 $.

The functions of the noncommutative basis of solutions (\ref{noncH}) are the eigenfunctions of the spin operator (\ref{spinH})  and are not eigenfunctions of the momentum operator $\hat p_2 = -iX_2$:
\begin{equation}\nonumber
	\hat S \psi_\sigma = \zeta \sqrt{e H}\psi_\sigma,\quad
	\hat p_2 \psi_\sigma = i l_2(q) \psi_\sigma.
\end{equation}
Thus, the dependence on the variables $(x,y)$ is expressed in terms of elementary functions. The reduced system (\ref{reductH}) is generally different from the reduced system of equations obtained by the noncommutative reduction method, unlike the case of a spherically symmetric field, where the  reduced system of equations is the same for the both  bases of solutions. Therefore, elucidating the connection between basis (\ref{solrazdH}) obtained by the method of separation of variables and basis  (\ref{noncH}) obtained by the noncommutative reduction is more difficult problem than in the case of a spherically symmetric potential, and is the subject of a separate study.

\section{Electromagnetic field contracting a subalgebra of symmetry operators}

Consider an electromagnetic field
\begin{equation}
	\mathbf{H} = [\mathbf{n},\mathbf{E}],\quad (\mathbf{n},\mathbf{E}) = 0,\quad
	\mathbf{n} = (0,0,-1),
	\quad E_x = \frac{\alpha}{t-z},\quad E_y = \varphi(y),
	\label{BGYfield}
\end{equation}
where $\alpha$ is a real parameter and  $\varphi(y)$ is a smooth function. The potential of the field (\ref{BGYfield}) can be taken as
\begin{equation}\nonumber
	A_2 = \alpha,\quad A_3 = - \frac{\alpha}{\varepsilon},\quad
	A_4 = - A_1 = \frac{\alpha x}{t-z} + \varphi(y),\quad \varepsilon > 0.
\end{equation}
The Dirac equation on $\mathcal{M}(t,x,y,z)$
\begin{equation}
	\hat H\psi = m\psi,\quad\hat H = \gamma^\mu\hat{\mathcal{P}}_\mu,\quad
	\hat{\mathbf{\mathcal P}}_\mu = \hat p_\mu - e A_\mu
	\label{BGYdirac}
\end{equation}
when $e=0$ admits the Lie algebra $\mathfrak{g}$ of symmetry operators 
\begin{gather}\nonumber
	\hat X_1 = \hat L_{21} + \hat L_{24} + \frac{i}{2}(\gamma^{12}+\gamma^{24}),\quad
	\hat X_2 = \hat L_{14} -\frac{i}{2}\gamma^{14},\quad
	\hat X_3 = \hat \partial_t + \partial_z,\quad
	\hat X_4 = \hat \partial_x + \varepsilon \hat \partial_y,
\end{gather}
with with non-trivial commutators
\begin{equation}\nonumber
	[\hat X_1,\hat X_2] = \hat X_1,\quad
	[\hat X_1,\hat X_4] = -\hat X_3,\quad
	[\hat X_2,\hat X_3] = -\hat X_3.
\end{equation}
Here $\hat L_{\mu\nu} = x_\mu\partial_\nu - x_\nu\partial_\mu$, $x_\mu = (t,-x,-y,-z)$.
In  presence of the external electromagnetic field (\ref {BGYfield}), the  Dirac equation admits  the  Abelian subalgebra $\{\hat X_1, \hat X_3 \} $ of an algebra $\mathfrak {g} $ as the algebra of differential symmetry operators of the first order. However, there are no spin symmetry operators.
That is, the inclusion of the electromagnetic field $(\ref{BGYfield})$ contracts the symmetry algebra of the free Dirac equation to the commutative subalgebra.

A necessary condition for the separation of variables in the Dirac equation is the existence of three commuting symmetry operators of the first order. In our case, this condition does not hold and for the reduction of the Dirac equation to a system of ordinary differential equations one needs  to go beyond the classical separation of variables. One way may be  the separation of variables in the squared Dirac equation $ (\hat H^2 - m^2)\phi = 0 $ with the use of a complete set of symmetry operators $ \{\hat Y_1, \hat Y_3, \hat p_ {22} \} $. Then the basis of solutions of  the original  Dirac equation (\ref{BGYdirac}) can be obtained in the form of $ \psi = (\hat H + m) \phi $.

We show that one can perform  the noncommutative reduction of the Dirac equation (\ref{BGYdirac}) in the external electromagnetic field (\ref {BGYfield}) using the Lie algebra $\mathfrak{g} $ of symmetry operators of the free Dirac equation.

Let $U$ be a neighborhood of a point with the coordinates $t=x=y=z=0$. Denote by $g=\{g_1, g_2, g_3, g_4\}$ a curvilinear coordinate system on $U$,
\begin{equation}
	t = -\frac{1}{2}(g_1^2 e^{g_2}+ e^{-g_2}),\quad
	x = g_4 - g_1,\quad
	y = \varepsilon g_4,\quad
	z = -\frac{1}{2}(g_1^2 e^{g_2} - e^{-g_2}).
	\label{sysc}
\end{equation}
The local Lie group $G=\exp(\mathfrak{g})$ acts simply transitively on Minkowski space. We can identify $U(g)$ with the local Lie group $G\simeq U(g)$ by introducing on $U(g)$ the group multiplication law:
\begin{gather}\nonumber
	(g\cdot g')_1 = g_1 + e^{-g_2}{g'}_1,\quad
	(g\cdot g')_2 = g_2 + {g'}_2,\\ \nonumber
	(g\cdot g')_3 = {g'}_3 + e^{{g'}_2}(g_3 + g_4 {g'}_1),\quad
	(g\cdot g')_4 = g_4 + {g'}_4.
\end{gather}
Note that (\ref{sysc}) is the canonical coordinate system of the second kind on the Lie group $G$. The left-invariant $\xi(g)$ and right-invariant $\eta(g)$ basis vector fields on the Lie group $G$ have the form
\begin{gather}\nonumber
	\xi_1 = e^{-g_2}\partial_1 + g_4\partial_3,\quad
	\xi_2 = \partial_2 + g_3\partial_3,\quad
	\xi_3 = \partial_3,\quad
	\xi_4 = \partial_4,\\ \nonumber
	\eta_1 = -\partial_1,\quad
	\eta_2 = g_1\partial_1-\partial_2,\quad
	\eta_3 = - e^{g_2}\partial_3,\quad
	\eta_4 = - (g_1 e^{g_2}\partial_3+\partial_4).
\end{gather}
The operators of $\lambda $-representation have the form
\begin{equation}\nonumber
	l_1 = i q_1 \exp(-q_2),\quad
	l_2 = \partial_{q_2},\quad
	l_3 = i \exp(-q_2),\quad
	l_4 = \partial_{q_1},\quad (q_1,q_2)\in Q = \mathbb{R}^2.
\end{equation}
We introduce the measure $d\mu (q) = d q_1 d q_2$ in the space $Q$ with respect to which the operators $ -il_a(q,\partial_q)$ are Hermitian. Solving the system of equations (\ref{sysXLg}) we obtain
\begin{equation}\nonumber
	D_{q q'}(g^{-1}) = \exp\left( - i\left[ g_3 e^{-g_1-{q'}_2} + g_1 {q'}_1 e^{-{q'}_2} \right] - 2{q'}_2 \right)\delta(g_4+{q'}_1-q_1)\delta(g_2 + {q'}_2 -q_2).
\end{equation}
Let us  write the Dirac equation in the new coordinate system. The metric in the curvilinear coordinate system (\ref{sysc}) is represented as a right-invariant metric on a Lie group $G$ with the metric tensor
\begin{equation}\nonumber
	g^{\mu\nu}(g) = G^{a b}\eta_a^\mu(g)\eta_b^\nu(g),\quad G^{a b} = (G_{a b})^{-1},\quad
	G_{a b} = \begin{pmatrix}
		-1 & 0 & 0 & 1\\
		0  & 0 & 1 & 0\\
		0  & 1 & 0 & 0\\
		1  & 0 & 0 & -(1+\varepsilon^2)
	\end{pmatrix}.
\end{equation}
Here, $G_{ab}$ are tetrad components of the metric tensor $g_{\mu\nu}$ with respect to the moving frame of the right-invariant vector fields $\eta_a(g)$. We decompose the Dirac gamma matrices $\gamma^\mu(g)$ in a curvilinear coordinate system $ g_\nu = g_\nu(t, x, y, z)$ in the moving frame: $\gamma^\mu (g) = \hat \gamma^a \eta_a^\mu (g)$. The constant gamma matrices $\hat \gamma^a $ are defined as an arbitrary but fixed solution of the equation $\{\hat\gamma^a, \hat \gamma^b \} = 2 G^{ab} $. Let us choose a solution of this equation in the form
\begin{equation}\nonumber
	\hat\gamma^1 = \frac{1}{\varepsilon}\gamma^3 - \gamma^4,\quad
	\hat\gamma^2 = -\frac{1}{2}(\gamma^1 - \gamma^2),\quad
	\hat\gamma^3 = -(\gamma^1 + \gamma^2),\quad
	\hat\gamma^4 = \frac{1}{\varepsilon}\gamma^3.
\end{equation}
The Dirac operator (\ref{BGYdirac}) takes the form
\begin{equation}
	\hat H = \hat\gamma^a \left(i (\eta_a(g)+\Gamma_a) - e A_a(g)\right),
	\label{HG}
\end{equation}
where $\Gamma_a$ are tetrad components of the spin connection
\begin{equation}\nonumber
	\Gamma_1 = -\frac{1}{2}(\hat\gamma^{12}+\hat\gamma^{24}),\quad
	\Gamma_2 = -\frac{1}{2}\hat\gamma^{23},\quad
	\Gamma_3 = \Gamma_4 = 0.
\end{equation}
The electromagnetic potential $A_a(g) = A_\mu(g) \eta^\mu_a(g)$ is given by the following tetrad components
\begin{equation}\nonumber
	A_1(g) = \alpha,\quad
	A_2(g) = \exp(-g_2)\phi(\varepsilon g_4) - \alpha g_4,\quad
	A_3(g) = A_4(g) = 0.
\end{equation}
We seek the basis of solutions for the Dirac equation with operator (\ref{HG}) in the form
\begin{gather}\label{noncanz}
	\psi_q(g) = \int_Q \hat \psi(q') D_{q q'}(g^{-1}) d\mu(q) =\\ \nonumber
 =	\exp\left( - i\left[ g_3 e^{-g_1-{q'}_2} + g_1 {q'}_1 e^{-{q'}_2} \right] - 2{q'}_2 \right)\hat\psi(q_1 - g_4, q_2 - g_2).
\end{gather}
Substituting (\ref{noncanz}) into the Dirac equation $\hat H \psi = m \psi $, we obtain the reduced Dirac equation with the Hamiltonian
\begin{gather}\nonumber
	\hat H = \hat\gamma^1\left[ i (l_1(q') + \Gamma_1) - e\alpha \right] +
	i\hat\gamma^3 l_3(q') + i\hat\gamma^4 l_4(q')+ \\ \nonumber
	+\hat\gamma^2\left[ i (l_2(q') + \Gamma_2) - e \left( \varphi(\varepsilon(q_1-{q'}_1)) e^{{q'}_2 - {q}_2} - \alpha(q_1 - {q_1}')\right) \right].
\end{gather}
Making a change of variables,  $u=q_1 - q_1^{'}, v =\exp(q_2^{'}-q_2)$, we can write the reduced Dirac operator as
\begin{gather}\label{redHQ2}
	\hat H =  -i\hat\gamma^4 \partial_u + i v \hat\gamma^2\partial_v - \frac{1}{v}e^{-q_2}\left( (q_1 - u)\hat\gamma^1 +\hat\gamma^3 \right)-e \alpha \hat\gamma^1 - \\ \nonumber 
	-e \hat\gamma^2\left( v \varphi(\varepsilon u) - \alpha u \right) + \frac{i}{2}\hat\gamma^2\left( \hat\gamma^1\hat\gamma^4 + \frac{1}{\varepsilon^2}+2 \right)
\end{gather}
The Dirac operator (\ref{redHQ2}) admits the following  symmetry operator of the first order:
\begin{equation}\nonumber
	\hat Y = -\partial_v + \frac{1}{2 v}\left(
	\hat\gamma^{23} + (u-q_1)(\hat\gamma^{12}+\gamma^{24}) + \frac{i}{v}e^{-q_2}(u-q_1)^2 + 2 i e \alpha q_1 (1-v)-1.
	\right)
\end{equation}
This the symmetry operator allows to perform separation of variables in the reduced Dirac equation. The solution of the equation $-i\hat Y \psi_\kappa(u,v) = \kappa \psi_\kappa(u,v)$ is
\begin{gather}\label{Ysol}
	\psi_\kappa(u,v) = \frac{1}{\sqrt{v}}\exp\bigg{(} \frac{1}{2}\left[ \hat\gamma^{23} + (u - q_1)(\hat\gamma^{12}+\hat\gamma^{24}) + 2 i e \alpha q_1 \right] \log v  - \\ \nonumber
	- i(\kappa + e \alpha q_1) v -\frac{i}{2 v}e^{-q_2} (u-q_1)^2\bigg{)}\Phi_\kappa(u).
\end{gather}
Substituting (\ref{Ysol}) into the Dirac equation $\hat H \psi_\kappa(u,v) = m \psi_\kappa(u,v) $, we leads to a system of ordinary differential equations:
\begin{gather}\label{ordnYH}
	-i\hat\gamma^4 {\Phi'}_\kappa(u) + \bigg{(} \hat\gamma^1 \left( (u-q_1)e^{-q_2} - e \alpha \right) -
	e^{-q_2} \left( \hat\gamma^3 + (u - q_1)\hat\gamma^4\right) + \\ \nonumber
	+\hat\gamma^2 \left( \frac{i}{2}\hat\gamma^1\hat\gamma^4 -\frac{1}{2}e^{-q_2}(u-q_1)^2 - e \varphi(\varepsilon u) + e \alpha u + \frac{1}{2\varepsilon^2} + \kappa \right) - m \bigg{)}{\Phi}_\kappa(u) = 0.
\end{gather}

\section{Conclusion}

We explored the integrability features of the Dirac equation with an external electromagnetic field by means of the noncommutative integration method and the method of separation of variables in terms of external fields of special form. Both methods use essentially the Lie algebra of symmetry operators of the Dirac equation. In the examples considered, we studied changes of the subalgebras of symmetry operators that are used to construct exact solutions of the Dirac equation with an external field, compared to the subalgebras of the free Dirac equation.

In the classical problem of integration of the Dirac equation with a spherically symmetric potential, a noncommutative reduction was carried out using the Lie algebra $\mathfrak{so}(3)$ of differential symmetry operators of the first order. It was shown that the reduced Dirac equation obtained by means of the NI method is identical to the similar equation, which occurs in the SoV method.This is due to the fact that in this problem the inclusion of an external field does not change the subalgebra $\mathfrak{so}(3)$ of the Lie symmetry algebra of the free Dirac equation, which is used for constructing solutions. The relationship between the NI-solutions and the separable solutions was found. The angular part of the NI-solution is expressed in terms of elementary functions and these solutions are of interest in the quantum theory of angular momentum. An expression was obtained for the spherical spinors (\ref{omega2}) which implies that a continuous parameter $ q $ numbering the basis functions of the NI-solutions is a dual value with respect to the projection of the momentum on the preferred direction.

By the example of the Dirac equation in an external electric $E_z = E_z (z)$ and a uniform static magnetic field $H = H_z$, we investigated the situation when an external field leads to a central extension of the Lie algebra of symmetry operators. In this case the reduced Dirac equation, obtained by the noncommutative integration method, is shown to be the same as the similar equation arising in the SoV method only in a special case.

The noncommutative reduction was performed for the Dirac equation in the crossed fields (\ref{BGYfield}). The external field contracts the Lie algebra of symmetry operators to its Abelian subalgebra. In this case the Dirac equation does not admit separation of variables. But, it is possible to perform the noncommutative reduction of the Dirac equation to a system of ordinary differential equations (\ref{ordnYH}) using a subalgebra of symmetry operators for the free Dirac equation.

\section*{Acknowledgments}

The work was partially supported by Tomsk State University Competitiveness Improvement Program and by program 'Nauka' under contract No. 1.676.2014/ K.



\begin{thebibliography}{22}

	\bibitem{Bagrov} Bagrov V G and Gitman D M 1990 {\it Exact solutions of relativistic wave equations} (Dordrecht: Kluwer);

	\bibitem{BagrovG} Bagrov V G and Gitman D M 2014 {\it The Dirac equation and its Solutions}(Boston: De Gruyter);

	\bibitem{Simulik} Simulik V M and Krivski I Yu 2014 Link between the relativistic canonical quantum mechanics and 
	the Dirac equation {\it Univ. J. Phys. Appl.} {\bf 2} 115

	\bibitem{Sitenko} Sitenko Yu A and Yushchenko S A 2014 The Casimir effect with quantized charged scalar matter in background magnetic field {\it Int. J. Mod. Phys. A } {\bf 29} 1450052

	\bibitem{FrGitShv}Fradkin E S,  Gitman D M and Shvartsman Sh M 1991 {\it Quantum Electrodynamics with Unstable Vacuum} (Springer-Verlag, Berlin Heidelberg New-York London Paris Hong-Kong Barcelona)

	\bibitem{SpSh1} Shapovalov A V and Shirokov I V 1995 Noncommutative integration of linear differential equations {\it Theor. Math. Phys.} {\bf 104} 921

	\bibitem{SpSh2} Shapovalov A V and Shirokov I V 1996 Noncommutative integration method for linear partial differential equations. Functional algebras and dimensional reduction  {\it Theor. Math. Phys.} {\bf 106} 1

	\bibitem{Br01} Breev A I and  Shapovalov A V 2014 Yang-Mills gauge fields conserving the symmetry algebra of the Dirac equation in a homogeneous space {\it Journal of Physics: Conference Series}, {\bf 563} 012004

	\bibitem{Br02} Breev A I and Shirokov I V 2009 Polarization of a spinor field vacuum on manifolds of the lie groups {\it Russian Physics Journal} {\bf 52} 823

	\bibitem{Klish1} Klishevich V V 2001 Exact solution of Dirac and Klein--Gordon--Fock equations in a curved space admitting a second Dirac operator {\it Class. Quantum Grav.} {\bf 18} 3735

	\bibitem{Klish2} Klishevich V Vand Tyumentsev V A 2005 On the solution of the Dirac equation in de Sitter space {\it Class. Quantum Grav.} {\bf 22} 4263

	\bibitem{Varaksin} Varaksin O L and Shirokov I V 1996 Integration of the Dirac equation, which does not
	presume complete separation of variables, in Stackel spaces {\it Russian Physics Journal} {\bf 39} 27

	\bibitem{ShVN} Shapovalov V N 1975 Symmetry of Dirac-Fock equation {\it Izv. Vuz. Phys.} \textbf{6} 57.

	\bibitem{Carter} Carter B and McLenaghan R G 1979 Generalized total angular momentum operator for the Dirac equation in curved space-time {\it Phys. Rev. D} \textbf{19} 1093.

	\bibitem{Magazev} Magazev A A 2012 Integration Klein-Gordon-Fock equations in an external electromagnetic field on Lie groups {\it Theor. Math. Physics} \textbf{3} 1654.

	\bibitem{Obukhov} Bagrov V G and Obukhov V V 1992 New method of integration for the Dirac equation on a curved space-time  {\it Journal of Mathematical Physics} \textbf{33} 2279.

	\bibitem{Akhiezer} Akhiezer A I and Berestetskii V B 1965 {\it Quantum Electrodynamics} (New York: Interscience Publishers).

	\bibitem{Edmonds} Edmonds A R 1957 {\it Angular Momentum in Quantum Mechanics} (Princeton: Princeton University Press)


	\bibitem{ShSO3} Baranovsky S P, Mikheev V V and Shirokov I V 2001 Quantum Hamiltonian Systems on K-Orbits: Semiclassical Spectrum of the Asymmetric Top {\it Theor. Math. Phys.} {\bf 129} 1311

	\bibitem{GnSO3} Gonchapovskii M M and Shirokov I V 2009 An integrable class of differential equations with nonlocal nonlinearity on Lie groups {\it Theor. Math. Phys.}, {\bf 161} 1604.

	\bibitem{Br14}Breev A I 2014 Scalar field vacuum polarization on homogeneous spaces with an invariant metric {\it Theor. Math. Phys.} {\bf 178} 59

	\bibitem{Kirr} Kirillov A A 1976 {\it Elements of the Theory of Representation} (New York: Springer)

\end{thebibliography}
\end{document}